# Power System Transient Modeling and Simulation using Integrated Circuit

Xiang Zhang[1], Member, IEEE, Renchang Dai[1], Senior Member, IEEE, Peng Wei[1], Student Member, IEEE, Yijing Liu[1], Student Member, IEEE, Guangyi Liu[1], Senior Member, IEEE, Zhiwei Wang[1], Senior Member, IEEE

[1]Global Energy Interconnection Research Institute, San Jose, CA, 95134, USA

*Abstract*—Transient stability analysis (TSA) plays an important role in power system analysis to investigate the stability of power system. Traditionally, transient stability analysis methods have been developed using time domain simulation by means of numerical integration method. In this paper, a new approach is proposed to model power systems as an integrated circuit and simulate the power system dynamic behavior by integrated circuit simulator as preliminary exploration for the ambition of using very large-scale integrated circuit chip to mimic power system behaviors. The proposed method modeled power grid, generator, governor, and exciter with high fidelity. The power system dynamic simulation accuracy and efficiency of the proposed approach are verified and demonstrated by case study on an IEEE standard system.

*Index Terms*-- Circuit Simulation, Equivalent Circuit Model, Terminal Circuit, Transient Stability, Very Large-scale Integrated Circuit

## I. INTRODUCTION

The modern power systems are evolving into large-scale and complex electric circuit systems with high penetration of power electronics equipment in both HVAC and HVDC systems [1]–[4]. Emerging technologies including ultra-high voltage hybrid AC-DC transmission systems, distributed renewable power, and energy storage systems required advanced and complex control strategies on the existing power grid [5]–[8]. To secure the power system operation, it is crucial to analyze the complex power system behavior not only in steady-state but in transient state.

Traditionally, mathematical model representing power system transient behavior is a large Differential-Algebraic Equations (DAE). Time domain simulation is usually applied to simulate power system transient. Transient stability analysis imposes heavy computation burden due to the large and complex algebraic and differential equations involved in transient model along with the nonlinear characteristics of power system components.

Physically, power system is a very large-scale circuit. The power system network consists of transmission lines and transformers which are consisted of basic circuit components including resistors, inductors, and capacitors. Using very large-scale integrated (VLSI) circuit to model power system with high fidelity is straightforward and promising which could mimic power system behaviors in real time supporting operators to be better aware of system steady-state and transient operation situation and provides researchers and engineers opportunity to investigate various phenomena in power systems which are possible overseen in the mathematical models. In industry practice, VLSI circuits are designed and simulated by electronic design automation (EDA) tools before they are launched to manufacture. Using VLSI circuit to model power system and simulate power system transient by circuit simulator is a niche approach which potentially enables engineer to manufacture a VLSI circuit chip to represent power system and replace numerical integration method to mimic power system steady-state and dynamic behaviors. Under this ambition, as a preliminary exploration, this paper proposes an approach to use integrated circuit to model power system and simulate power system transient.

Researchers have been proposed an approach to solve the nonlinear power flow formulation using a terminal circuit model [9]. This circuit analysis eliminates the nonlinear term in the Jacobian matrix introduced by the components such as generator and inconstant impedance loads. Instead, the current injections by such components are modeled by a constant submatrix module and implemented to the current balance equation in an attempt to keep the Jacobian matrix constant [9], [10].

In this paper, the terminal circuit method is further developed and applied to transient simulation. In this method, the dynamic models of electric machines are encapsulated in the circuits and interfaced with power grid by current injection of the terminal circuits satisfying the Kirchhoff's Voltage Law (KVL) and the Kirchhoff's Current Law (KCL) [9]–[11]. Using this method, the current injection matrix and the respective Jacobian matrix only requires to be updated when the power network topology is changed. Subsequently, the power flow and the respective dynamic transient models can be created independently and calculated alternatively.

In this paper, the proposed modeling approach and detailed circuit models for power system simulation are discussed and illustrated. In addition, the dynamic modeling of a 6[th] order

This work is supported by the State Grid Corporation technology project 5455HJ180021.

synchronous machine with an associated exciter and a governor is represented by SPICE circuit models. In the section IV, the proposed method and approach are tested on the IEEE-14 bus system. The computation accuracy and efficiency of the transient simulation calculation using a circuit simulator is verified by comparing the results against the commonly used commercial tool – PowerWorld Simulator.

## II. BASIC CIRCUIT COMPONENTS

The bus-branch model of a power system network consists of nodes, transmission lines, transformers, and power loads, etc. These system components are physically resistors, inductors, capacitors, voltage source, or their combinations which can be modeled into integrated circuit. The steady-state bus-branch model of a power system has been applied in [9] and [12]. To further extend the approach to transient simulation, the dynamic model of a generator is constructed using the circuit model along with exciters, governors, stabilizers and other control systems. The transfer functions of the control systems typically involve basic circuit blocks, such as adder, limiter, multiplier, integrator, low pass filters, high pass filters, and lead-lag filters, etc. These control blocks can be built by circuits in SPICE-level language as discussed as follow.

For instance, the transfer function diagrams of low pass filter and high pass are shown in Fig. 1(a) and Fig. 1(c). Meanwhile, the two filters are constructed using ideal circuit elements as shown in Fig. 1(b) and Fig. 1(d) in integrated circuit. By carefully setting the parameters, the RC circuits are functionally the same as the filters.

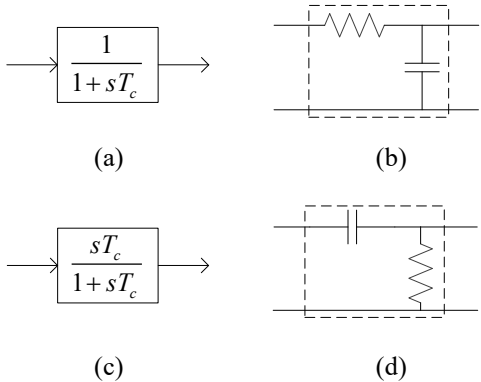

Fig. 1. Low and high pass filter block diagrams and circuits

Nonlinear dependent voltage and current source, shown in Fig. 2, are also known as B-source devices and commonly used as analog behavioral modeling (ABM) in integrated circuit[13]. They are adopted to accommodate non-ideal transformer and other complex characteristics in power system integrated circuit modeling. The nonlinear B-source devices in Fig. 2 can model complex relations in power system by defining the expression functions. By which, the input signal(s) of a nonlinear B-source device is dependent or a function of state variables of power systems.

In addition, commonly used operator blocks shown in Fig. 3 such as gain, adder, and product are realized in circuit simulator using a nonlinear dependent voltage B-source device shown in Fig. 2(a). For example, the multiple input signals $V_{i,1}$ to $V_{i,N}$ are fed into the nonlinear dependent voltage source in order to model multi-input adder and product as shown in Fig. 3(b)-(c). The output expression for the three operator blocks are shown in (1)-(3), where $K_n$ in (2) are the coefficients for the operator. The subtractor and divider can be derived from (1) and (2) by changing the operating signs.

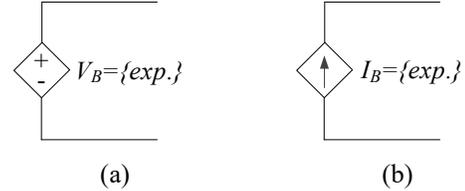

Fig. 2. Nonlinear dependent source circuits

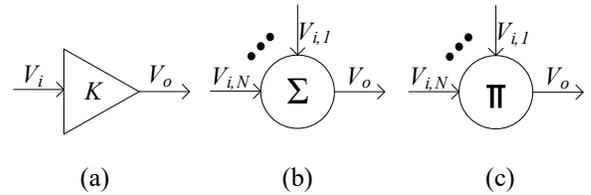

Fig. 3. Block diagrams for operators

$$V_o = K \cdot V_i \qquad (1)$$
$$V_o = \sum_{n=1}^{N} K_n \cdot V_{i,n} \qquad (2)$$
$$V_o = \prod_{n=1}^{N} V_{i,n} \qquad (3)$$

With the help of the commonly used blocks, a lead-lag filter is constructed using a combination of low pass, high pass filter, and an adder. The block diagram and the detailed construction are shown in Fig. 4. The time constants and gains in the channels $V_L$ and $V_H$ are constructed to function the lead-lag filter shown in Fig. 4(a).

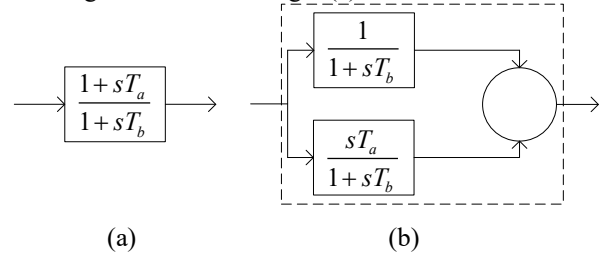

Fig. 4. Lead-lag filter block diagram and circuit

Integrator is an important and commonly used block in control systems. Its performance takes a great role in improving the efficiency of integrated circuits. The function block diagram of such an integrator is shown Fig. 5(a). Meanwhile, the time domain integration algorithm is replaced by a circuit as shown in Fig. 5(b). The integrator circuit is a nonlinear dependent current source in parallel with an ideal capacitor as depicted in Fig. 2(b). Utilizing the capacitor voltage update equation stated in (4), the nonlinear dependent current source provides a time dependent current $f(V_i)$ that simulates the time derivative of a state variable labeled as $V_o$.

The capacitance of the capacitor is set as $\tau$ to make the state variable $V_o$ representing the desired integration.

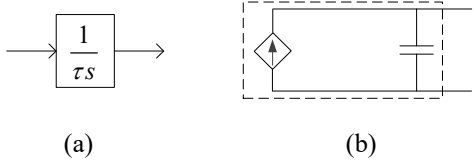

Fig. 5. Integrator block diagram and circuit

$$\frac{dV_o}{dt} = \frac{1}{C} f(V_i) \qquad (4)$$

III. GENERATOR AND CONTROL SYSTEM DYNAMIC MODEL

Typically, a power network consists of basic circuit elements such as resistors, inductors, and capacitors, and generators. The current balance network matrix is formed using the Kirchhoff's Current Law upon the network topology. To fit into the circuit simulator, the complex number operation is dissembled into real and imaginary part. The power flow equation is in the format of YV=I [9], [12], [13]. In power grid, the transmission line is well-established and modeled as a π-type equivalent circuit with parameters of line resistance (R), line reactance (X), and shunt susceptance (B)[12][14] as shown in Fig. 6. The transformer is modeled using a coil resistance (R), a coil reactance (X), and a turns ratio (n)[12][14] as shown in Fig. 7, where the $Y=(R+jX)^{-1}$. Using this formation, the ideal transformer tap change is integrated into the three components of the π shape. They can be modeled by resistors, inductors, and capacitors along with loads straightforwardly in integrated circuit model.

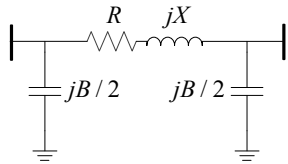

Fig. 6. Equivalent π circuit for transmission line

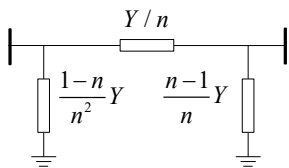

Fig. 7. Equivalent π circuit for transformer

In addition to the previously defined power grid components transmission line and transformer, constant power/impedance load and generator along with its control system dynamic model are needed for the transient simulation using the basic circuit models depicted in Section II.

A. Synchronous Machine Modeling

A synchronous machine has three-phase stator windings, field windings, and three damper windings. To comply with the positive-negative-zero(α-γ-0) sequence of the power system, the machine dynamics are transformed into the Direct-Quadrature-Zero(DQZ) coordinate using the Park's transformation along with the per-unitized system [2], [15]–[17]. A sixth-order synchronous machine model is adopted to construct the circuit. In a circuit simulator, the time derivative terms of machine dynamic equations for excitation voltage and flux linkage are achieved using the circuit based integrator derived in Fig. 5. The right-hand sides of the integral equations provide the time based nonlinear current source while the state variables are evolutions of the ideal capacitor voltage. In addition, the time constant of the integral equations are the capacitance of ideal capacitor. The algebraic equations are modeled using the nonlinear dependent voltage and current sources shown in Fig. 2.

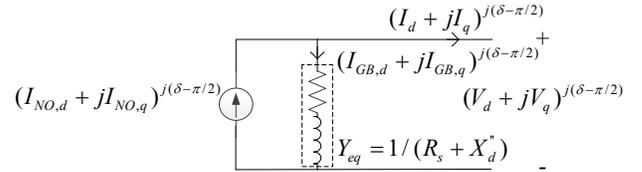

Fig. 8. Norton equivalent circuit in the network-based reference frame

Generator is interfacing with power grid by the terminal voltage and injection current. To formulate a current injection, a Norton equivalent circuit in Fig. 8 is derived from the Thevenin equivalent circuit of a generator proposed from [2] [11][17]. In Fig. 8, the $I_{NO,d}$ and $I_{NO,q}$ and $I_d$ and $I_q$ are the Norton DQZ reference current and the α-γ-0 reference current sending back to the network. $R_s$ and $X_d''$ are the resistance and d-axis flux sub-transient reactance. In addition, δ is the power angle.

B. Exciter Modeling

The IEEE Type 1 Excitation System[18], [19] is taken as an example to illustrate the approach to modeling excitation system by circuit to regulate the generator terminal voltage magnitude $V_t$ as shown in Fig. 9.

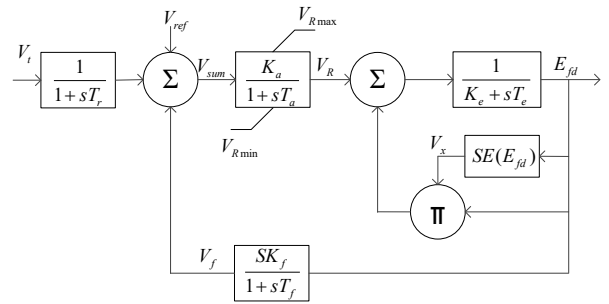

Fig. 9. Exciter IEEET1 control block diagram

The control blocks in Fig. 9 are modeled by the circuits of low pass filter, high pass filter, the gain blocks, or their combinations developed in Fig. 1 and Fig. 3. $SE(E_{fd})$ is a saturation function of $E_{fd}$ which is modeled using a nonlinear dependent voltage source in Fig. 2(a). $V_{ref}$ is the known reference voltage. $V_{Rmax}$ and $V_{Rmin}$ are the upper and the lower limits of the voltage regulator whose model is also developed via a nonlinear dependent voltage source in Fig. 2(a).

## C. Governor Modeling

WSCC Type G model [20], [21] shown in Fig. 10 is modeled by circuit as an example to accommodate the IEEE 14-bus system case study. In Fig. 10, *K* and *F* are the gains on frequency difference $\Delta\omega$ in (5) and the filtered signal. $T_n$ (*n*=1,2,…5) are the time constants. $P_{max}$ is the upper limit of the regulator. $P_{Mref}$ is the mechanical power command. $P_m$ is the mechanical power output. This model is constructed using 2 sets of lead-lag filter circuits, 2 sets of low pass filter circuits, one adder circuit, and a nonlinear dependent voltage source.

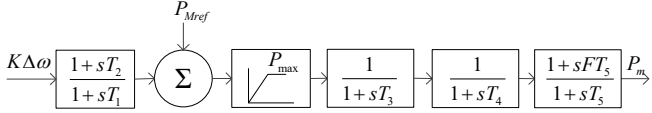

Fig. 10. Governor BPA GG control block diagram

$$\Delta\omega = \omega_s - \omega_e \quad (5)$$

## D. Constant Impedance/Power Load Modeling

In an integrated circuit, a constant impedance load is simply modeled as RLC circuit. A constant power load is modeled as a nonlinear dependent current source. The current is expressed as:

$$\tilde{I} = I_r + jI_i = \left(\frac{\tilde{S}_0}{\tilde{V}}\right)^* = \frac{(P_0 V_r + Q_0 V_i) + j(P_0 V_i - Q_0 V_r)}{|\tilde{V}|^2} \quad (6)$$

where $S_0$, $P_0$, and $Q_0$ are the given constant complex power, real power, and reactive power, respectively. The notation '~' on the *V*, *I*, and $S_0$ represents complex numbers. A voltage-in current-out model can be derived from (6) using the nonlinear dependent current source shown in Fig. 2 to model the constant power impedance.

## IV. CASE STUDY

To validate and demonstrate the accuracy of the proposed method, a case study on a modified IEEE 14-bus system is employed whose one-line diagram are shown in Fig. 11[22]. The parameters for the bus-branch model are acquired from the standard IEEE-14 bus system which can be found in [22] and the dynamic machine parameters are acquired and modified from [3], [23]. In this system, two generators and three condensers are involved where the voltage of the condensers is regulated by IEEET1 exciters. The dual transmission lines between Bus 1 and 2 are merged into one transmission line in the modified systems [3], [23]. The tap ratio of transformers between Bus 7 and Bus 8, and Bus7 and Bus 9 is 1.0[22], [23]. In summary, this system includes 14 buses, 3 transformers, and 17 transmission lines.

To demonstrate the computation accuracy and efficiency of the transient simulation calculation using integrated circuit simulator compared against PowerWorld Simulator, a three-phase to ground short circuit fault through a 0.2p.u. inductor at the middle point of the line between Bus 2 and Bus 4 is applied at t=1s. The fault is cleared at t=2s. The fault is realized using a circuit switch of which one terminal is attached to the middle point of the branch and the other terminal is connected to the ground.

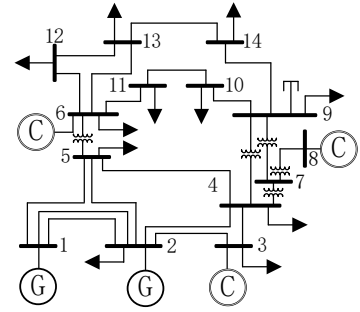

Fig. 11. One-line diagram of the IEEE 14-bus system[22]

The bus voltage magnitude at the generator and condenser terminal bus is shown by solid line in Fig. 12 and the speed ω and power angle δ of generators and condensers are shown by solid lines in Fig. 13 and Fig. 14, respectively. The simulation results are compared against that by PowerWorld Simulator depicted by dashed line. The simulation was conducted using an Intel Core i7-6600U @2.6GHz processor. For this case, it takes 2.78s and 3.20s to simulate 20 seconds dynamics by the integrated circuit model simulator and the PowerWorld Simulator, respectively.

Shown in Fig. 12 and Fig. 14, it is observed that the bus voltages drop during the short circuit fault and recover after fault is cleared. When the voltage drop, generation power could not be fully transmitted to power grid. The unbalanced mechanical power and electrical power accelerated generator speed. It takes several seconds after the fault is cleared that the system settled down to an equilibrium point.

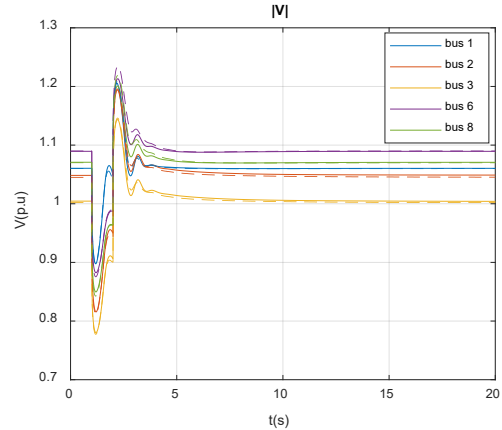

Fig. 12. Voltage magnitude for generator/condenser buses

Comparing with the simulation results by using circuit simulator and PowerWorld Simulator shown in Fig. 12 and Fig. 14, the dynamic behavior of the system before, during, and after disturbance are consistent in terms of bus voltage, generator speed, and power angle that justified the proposed power system transient simulation model and approach. The mismatch between the two solutions are mainly due to the model difference utilized by the proposed circuit simulator and Powerworld. The detailed models are not exactly same although some of the components are explicitly expressed in the Powerworld documents. In addition, different integration method and nonlinear solver are adopted by the two simulation approaches. Specifically, the power angle is essentially an integration of rotor speed, whose dynamic performance is not the same, although trivial to observe. As a result, this discrepancy is enlarged in Fig. 14.

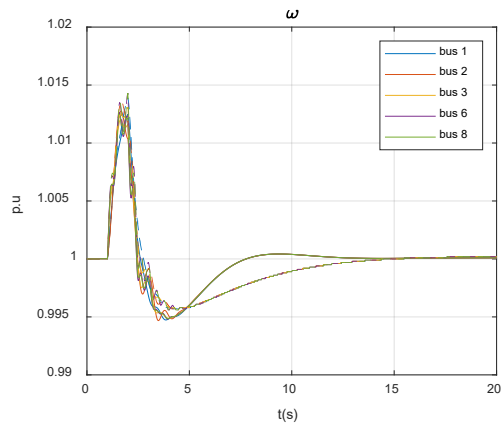

Fig. 13. Rotor speed of generators/condensers

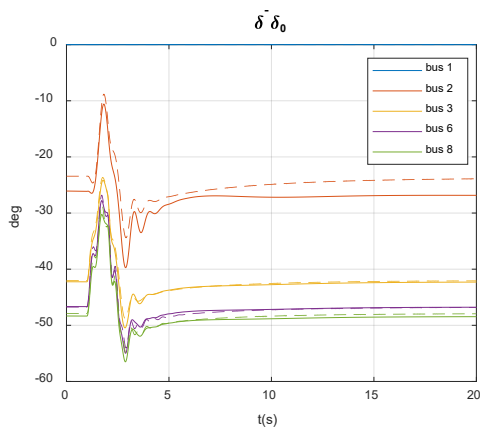

Fig. 14. Generator/condenser power angles referenced to bus 1

## V. Conclusion

This paper presents a circuit model for the transient simulation of a complex power system in conjunction with the dynamics of the electric machines and control systems. To investigate the transient stability by integrated circuit, power grid and a $6^{th}$ order sub-transient dynamic machine model has been developed by SPICE-level analog circuit language. The simulation accuracy of the test case proofed the concept of using integrated circuit to simulate power system behavior.

This paper is a concept proofing of using circuit simulator for transient simulation and the power system modeling in circuit simulator which is a fundamental work to enable a chip eventually to mimic power system and measure power system dynamics on the circuit directly in real time other than calculated by numerical approach. As an initial practice, the proposed approach using integrated circuit simulators for power system simulation can be extended to simulate power system electromagnetic transient.